\documentclass[twocolumn,twoside,fleqn,showpacs,showkeys,pra,aps,10pt,superscriptaddress,final]{article}

\usepackage{hep}
\graphicspath{{figs/}}      %定义图的存放路径

\def\refname{References}

\def\journalname{??}

\def\@pacs@name{PACS numbers: }%
\def\@keys@name{Keywords: }%

\def\Dated@name{Dated: }%
\def\Received@name{Received }%
\def\Revised@name{Revised }%
\def\Accepted@name{Accepted }%
\def\Published@name{Published }%
\def\address{\replace@command\address\affiliation}%
\def\altaddress{\replace@command\altaddress\altaffiliation}%

\definecolor{orangec}{cmyk}{.24,.91,.96,.18}
\definecolor{orangecc}{cmyk}{.24,.94,.96,.18}
\definecolor{oorangec}{cmyk}{.8,.2,.5,.4}
\definecolor{ooorangec}{cmyk}{1,.9,0.08,.04}      %四色

\definecolor{orangec}{cmyk}{.15,.7,.96,.0}
\definecolor{orangecc}{cmyk}{.15,.7,.96,.0}
\newcommand{\ODDBOX}{\noindent\raisebox{-13.3cm}{\includegraphics{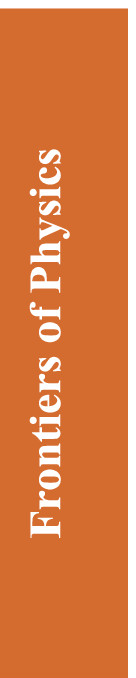}}}
\newcommand{\EVENBOX}{\noindent\raisebox{-13.3cm}{\includegraphics{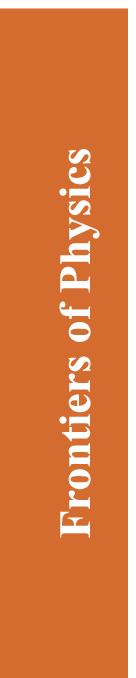}}}

\newfont{\yihao}{cmb10 at 18pt}
\newcommand{\Yihao}{\fontsize{18pt}{13.5pt}\selectfont}
\newcommand{\xiaosi}{\fontsize{11.5pt}{13.5pt}\selectfont}         % 小四字体   %\陈清华增加

\newfont{\xbt}{cmb10 at 12pt}

\def\frontmatter@title@format{%
    \centering%
    \usefont{T1}{fradmcn}{m}{n}\yihao}%
\def\@keys@name{{\color{ooorangec}\bf Keywords~~}}%
\def\@pacs@name{{\color{ooorangec}\bf PACS numbers~~}\vspace{2mm}}%
\def\frontmatter@authorformat{\vspace{5mm}\centering\bf}%

\newcommand{\catchline}[2]{
	{\vspace*{-16.4mm}\small%
	\noindent #1\\%
	\noindent #2\\[-2mm]%
    {\color{orangec}{\rule{\textwidth}{.5pt}}}\\[2mm]
    {\color{orangec}{\Yihao\textbf{\textsc{\Papertype}}}}\\[6mm]
    }\relax\par
}

\newcommand{\doi}[1]{DOI {#1}}
\renewcommand{\title}[1]
{\vspace*{-5mm}\begin{center}
{\Yihao\bf #1}
\end{center}
}

\renewcommand{\author}[1]
{\vspace*{0mm}
\begin{center}
{\bf #1}
\end{center}
}
\newcommand{\add}[1]{\begin{center}{\small\it #1}\end{center}}

\newcommand{\abs}[1]{
\begin{center}
\parbox[t]{156mm}{\noindent\color{oorangec}#1}
\end{center}}

\newcommand{\keywords}[1]{
\begin{center}
\parbox[t]{156mm}{\noindent{\bf\color{ooorangec}Keywords}\ \ #1}
\end{center}}

\newcommand{\pacsnumbers}[1]{
\begin{center}
\parbox[t]{156mm}{\noindent{\bf\color{ooorangec}PACS numbers}\ \ #1\vspace*{5mm}}
\end{center}}

\newcommand{\acknowledgements}[1]{\vspace*{4mm}\noindent{\renewcommand{\baselinestretch}{1.05}\footnotesize{\color{ooorangec}\bf Acknowledgements}\quad{#1}}}

\def\journalname{??}
\def\volumenumber#1{\gdef\@volumenumber{#1}}%
\def\@volumenumber{}%
\def\issuenumber#1{\gdef\@issuenumber{#1}}%
\def\@issuenumber{}%
\def\volumeyear#1{\gdef\@volumeyear{#1}}%
\def\@volumeyear{}%

\renewcommand\thesection{\arabic{section}}
\renewcommand\thesubsection{\arabic{section}.\arabic{subsection}}
\renewcommand\thesubsubsection{\arabic{section}.\arabic{subsection}.\arabic{subsubsection}}

\titleformat{\section}[hang]{\color{ooorangec}\vspace*{-1.2mm}\titlerule\vspace{1mm}\large\usefont{T1}{fradmcn}{m}{n}\xbt}{\thesection}{1em}{}
\titlespacing{\section}{0mm}{8mm}{5mm}

\titleformat{\subsection}{\normalfont\normalsize\color{ooorangec}}{\thesubsection}{1em}{}
\titlespacing{\subsection}{0mm}{5mm}{3mm}

\titleformat{\subsubsection}{\normalfont\normalsize\it\color{ooorangec}}{\thesubsubsection}{1em}{}
\titlespacing{\subsubsection}{0mm}{3mm}{3mm}

\titlecontents{section}
[0em] {\normalfont} {\thecontentslabel\quad} {\thecontentslabel}
{\titlerule*[.7pc]{}\contentspage}

\titlecontents{subsection}
[1.5em] {\normalfont} {\thecontentslabel\quad} {\thecontentslabel}
{\titlerule*[.7pc]{}\contentspage}

\titlecontents{subsubsection}
[4em] {\normalfont} {\thecontentslabel\quad} {\thecontentslabel}
{\titlerule*[.7pc]{}\contentspage}

\usepackage[small]{caption2}%去掉图注后昌号%%%[hang,footnotesize]表示图注在图号后自控对齐
\newlength{\halfpagewidth}
\divide\halfpagewidth by 1

\usepackage{mathrsfs}
\usepackage{ulem}

\begin{document}

\newcommand{\Papertype}{\sc Research article} % 文章类型
\def\volumeyear{2020} % 出版年
\def\volumenumber{15(1)} % 卷
\def\issuenumber{12XXXX} % 编号
\def\journalname{Front. Phys.} %期刊名
\newcommand{\doiurl}{10.1007/s11467-019-0935-y} % DOI链接
\newcommand{\allauthors}{Baijiong Lin, Xiangru Li and Woliang Yu} % 作者名称

\twocolumn[
\begin{@twocolumnfalse}
\catchline{\journalname~\volumenumber,~\issuenumber~(\volumeyear)}{\doi{\doiurl}} %Please ignore
\thispagestyle{firstpage}

%\begin{strip}

\title{Binary Neutron Stars Gravitational Wave Detection Based on Wavelet Packet Analysis And Convolutional Neural Networks}

\author{Baijiong Lin, Xiangru Li$^{*}$, Woliang Yu}

\add{South China Normal University, Guangzhou 510631, China\\
Corresponding author.\ E-mail: $^*$xiangru.li@gmail.com \\
Received August 15, 2019; Accepted October 02, 2019}

\abs{This work investigates the detection of binary neutron stars gravitational wave based on convolutional neural network (CNN). To promote the detection performance and efficiency, we proposed a scheme based on wavelet packet (WP) decomposition and CNN. The WP decomposition is a time-frequency method and can enhance the discriminant features between gravitational wave signal and noise before detection. The CNN conducts the gravitational wave detection by learning a function mapping relation from the data under being processed to the space of detection results. This function-mapping-relation style detection scheme can detection efficiency significantly. In this work, instrument effects are considered, and the noise are computed from a power spectral density (PSD) equivalent to the Advanced LIGO design sensitivity. The quantitative evaluations and comparisons with the state-of-art method matched filtering show the excellent performances for BNS gravitational wave detection. On efficiency, the current experiments show that this WP-CNN-based scheme is more than 960 times faster than the matched filtering.}

\keywords{Gravitational waves, Algorithms, Astrostatistics techniques}

\pacsnumbers{04.30.Db, 07.05.Mh}

\vspace*{-6mm}

\end{@twocolumnfalse}
]

\section{Introduction} \label{sec:intro}

\noindent The era of gravitational wave astronomy began with the binary black hole (BBH) merger GW150914 \cite{abbott2016observation} detected by advanced Laser Interferometer Gravitational-wave Observatory (LIGO \cite{aasi2015advanced}). Since then, nine additional BBH mergers have been observed \cite{abbott2016binary,abbott2016gw151226, scientific2017gw170104,abbott2017gw170608,abbott2017gw170814,ligo2018gwtc,ligo2018binary}. More special is that, on August 17, 2017, advanced LIGO and Virgo \cite{acernese2014advanced} together observed new gravitational wave from the inspiral of binary neutron stars (BNS), which is an entirely new type of gravitational wave source \cite{abbott2017gw170817}. The inspiral of two neutron stars, known as GW170817, was also seen by 70 observatories on seven continents and in space, across the electromagnetic spectrum, marking a significant breakthrough for multi-messenger astronomy \cite{abbott2017multi,metzger2017welcome,zhang2019delay}.

Matched filtering is the standard technique for gravitational wave data processing \cite{allen2012findchirp,González2013Gravitational}. This technique uses a  template bank consisting of a series of gravitational wave signals computed using  post-Newtonian theory \cite{blanchet2014gravitational,buonanno2009comparison}, the effective-one-body formalism \cite{pan2011inspiral,barausse2010improved,damour2016effective}, and numerical relativity \cite{chu2016accuracy,husa2016frequency,mroue2013catalog}. For detecting gravitational wave signal and estimating its parameters, matched filtering is to compute the similarities between every template and the signal under processing and find the one with the largest similarity. And to be accurate, the template bank should computed based on a dense gird over a parameter space, for example, masses, spins, inclination, sky location, orbit orientation, etc. Therefore, the bottleneck of matched filtering is its the computational burden.

In recent years, deep learning has attracted much attention in gravitational wave astronomy community and was widely investigated, for example, glitch classification \cite{george2017deep,razzano2018image,bahaadini2017deep}, gravitational wave detection \cite{gabbard2018matching, george2018deep,li2017method,gebhard2019convolutional}, signal denoise \cite{wei2019gravitational,torres2016denoising,shen2017denoising}, and wave source parameter estimation \cite{george2018deep,bouffanais2018bayesian}. In the specific context of gravitational wave detection, recent efforts have focused on the detection of the burst-like BBH gravitational wave signal  using convolutional neural networks (CNN) \cite{gabbard2018matching,george2018deep,li2017method,gebhard2019convolutional}.

However, there exist evident differences between BBH gravitational wave signal and binary neutron stars gravitational wave. Compared with the BBH systems, the BNS systems have smaller masses and longer durations. Besides, the BNS signals are almost only dominated by the inspiral phase, for the merger and post-merger phase happen at high frequencies, where advanced LIGO and Virgo are less sensitive \cite{ligo2018gwtc}. As a result, BNS signals are weaker, and  more challenging to be detected than BBH gravitational wave.

Therefore, this work studies BNS gravitational wave detection problem and proposes a detection scheme based on CNN and wavelet packet (WP) transform. The proposed scheme is evaluated using theoretical data on detection performance and efficiency with excellent results in comparing with the state-of-art method matched filtering.

\section{Experimental Data}\label{sec:data}

\noindent The gravitation wave (GW) signal is acquired after a long journey in space and a traversing in our instruments. Therefore, a piece of obtained data consists of noises, detector effects and GW signal. Furthermore, the observed GW signal is weak and buried in noises.

Therefore, the detectors should be very sensitive. However, the sensitivity results in our observations are affected seriously by random \& high-amplitude disturbances from environments and transient events. This kind interferences are referred to glitches in literature, and cause false positives. To reduce this kind positives, a scheme of  multiple detector-based vetoes are utilized in application \cite{ligo2018gwtc}. And there exists some sensitivity imbalances between different detectors, for example, Virgo detector, and the two LIGO detectors \cite{ligo2018gwtc}. This work investigated the situation with two detectors of LIGO, H1 and L1. To be clear, data from these two detectors are called H1 data set and L1 data set respectively.

The proposed scheme is of a supervised machine learning method. This kind methods always divides data sets into three parts: a training set, a validation set, and a test set. The training set represents our BNS gravitational wave detection problem and used for finding the optimal configurations of the proposed scheme; The validation set is used to tune the hyperparameters of model; And the test set is used to evaluate the performance of the proposed scheme \cite{bishop2006pattern}. Therefore, H1 data are divided into three subsets, and similarly for L1 data.

There are two kinds of samples in the training set, validation set and test set: one is pure noise, the other is gravitational wave signal contaminated with noise. To generated the experimental data of this work, therefore, two kinds of data components are needed: pure-noise and noise-free gravitational wave signal. These two kinds of data components are computed using the PyCBC (a Python-based toolkit for compact binary coalescence signal analysis) library  \cite{alex_nitz_2017_344823,usman2016pycbc,allen2012findchirp,allen2005chi,nitz2017detecting,dal2014implementing}. Actually, the noise is generated based on power spectral density (PSD) at the `Zero-detuned High Power' design sensitivity to reproduce the  sensitivity of LIGO detectors. The BNS gravitational wave signals are computed using SEOBNRv4T-type \cite{hinderer2016effects} waveform, which models the inspiral and merger phase for BNS gravitational wave signals \cite{abbott2018properties,ligo2018gwtc}.

An observed time-domain strain $s(t)$ can be formulated as follows:
\begin{equation}
	\label{eq:hn}
	s(t) = k\times h(t)+n(t),
\end{equation}
where $h(t)$ is a noise-free signal, $n(t)$ a  noise component, and $k$ is a proportional coefficient characterizing the relative intensity between noise and signal. If $k=0$, $s(t)$ is pure noise.

In simulation, we can adjust the value of $k$ to generate a signal data with a predefined optimal matched-filter signal-to-noise ratio (SNR \cite{owen1999matched}) defined as
\begin{equation}\label{Equ:rho:SNR}
	\rho = \frac{<s,u>}{\sqrt{<u,u>}},
\end{equation}
where $s$ is an observed strain, $u$ is a template waveform used to filter $s$. And the inner product
\begin{equation}
	<a,b>=4~Re[\int_0^{\infty}df\frac{\widetilde{a}^{*}(f)\widetilde{b}(f)}{S_n(f)}]
\end{equation}
is the noise-weighted cross-correlation between a and b. Here $\widetilde{a}(f)$ is the Fourier transform of a time-domain waveform $a(t)$ and the asterisk denotes a complex conjugate, and $S_n(f)$ is the PSD of the instrumental noise, which represents the sensitivity of the detector.

The BNS systems are simulated without spin and inclination. Every BNS system consists of two neutron stars, and suppose their masses are $m_1$ and $m_2$ respectively. For training data and validation data, the ranges of $m_1$ and $m_2$ are respectively from 1.36 to $2.26~M_{\odot}$, and from 0.86 to $1.36~M_{\odot}$, which are similar to GW170817 \cite{abbott2017gw170817}. While on the test set, $m_1$ and $m_2$ share a range from 1 to $3~M_{\odot}$ with constraint $m_1> m_2$. Therefore, some of the test data cover the parameter ranges of training data, while the parameters of the other test data are out of that of training data. The former subset of the test set is to evaluate the detection performance in an interpolating manner, and the latter in an extrapolating manner. These two subsets of test data shows the generalization performance of the proposed scheme from different aspects, more about the two kind generalization examinations are presented in Fig. \ref{fig:inside_outside_roc} and the fifth paragraph of section \ref{sec:experiments}. The simulated signals were truncated to 4 seconds in duration sampled at 16384 Hz, which retained the post-inspiral (and merger) phase of the signals. At the same time, we make the coalescence time of each signal appear randomly in the 4 seconds time series.

In application, it is impossible to guaranteed that an observation to be processed share the same parameters with anyone historical observation or theoretical template. Therefore, this work independently generated three template banks of noise-free BNS gravitational wave signals, Template\_Bank$_\texttt{training}$, Template\_Bank$_\texttt{validation}$ and Template\_Bank$_\texttt{test}$, used for computing training set, validation set and test set, respectively. The Template\_Bank$_\texttt{training}$,  Template\_Bank$_\texttt{validation}$ and Template\_Bank$_\texttt{test}$ consist of 4500 templates, 500 templates and 7000 templates respectively.

For detector H1, the corresponding training set, validation set and test set are respectively computed from template banks Template\_Bank$_\texttt{training}$,  Template\_Bank$_\texttt{validation}$ and  Template\_Bank$_\texttt{test}$, referred to TrH1, VaH1 and TeH1. Each template from  Template\_Bank$_\texttt{validation}$ is injected into a randomly generated noise strain to produce a sample for VaH1. This kind sample is named as signal+noise. In VaH1, another kind of sample is pure noise, and the ratio between the number of ``signal+noise" sample and pure noise sample is 1:1. Therefore, the VaH1 consists of 1000 samples. The test set TeH1 is generated similarly, and has 14000 samples. While, every template from  Template\_Bank$_\texttt{training}$ is injected into two independently generated noise and forms two ``signal+noise" samples for ``TrH1". This procedure, referred to ``data augmentation" in machine learning community, can not only increase the size of training set, but also avoid overfitting. In ``TrH1", the ratio  between the number of ``signal+noise" sample and pure noise sample is also 1:1. Therefore, the size of TrH1 is 18000. For detector L1, three data sets, TrL1, VaL1 and TeL1, are similarly computed from template banks Template\_Bank$_\texttt{training}$, Template\_Bank$_\texttt{validation}$ and Template\_Bank$_\texttt{test}$ respectively.

\section{Deep detection scheme}

\begin{figure}[!htp]
	\centering
	\includegraphics[width=0.5\textwidth]{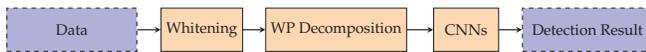}
	\caption{A flowchart of the proposed scheme. In this flowchart, the `Data' is an observation of time series $s(t)$, the input into `WP decomposition' is still a time series $s_w(t)$, and output from `WP decomposition' is a time-frequency map, which is input into the `CNNs'.}
	\label{fig:flowchart}
\end{figure}

\noindent The proposed scheme consists of three procedures: whiten the data under being dealt with; decompose the whitened data using wavelet packet into time-frequency maps; detect the existence of gravitational waves from binary neutron stars using two CNNs. A flowchart of the scheme is presented in Fig. \ref{fig:flowchart}.

\textbf{Whitening~~~~} In application, the obtained gravitational wave signals are buried in disturbances. Some components of the disturbances are non-stationary, and the non-stationariness can decrease the performance of our detection methods. Therefore, the first procedure in gravitation wave detection is whitening. The objective of a whitening procedure is to remove the correlation of the disturbances \cite{cuoco2004whitening,cuoco2001noise}.

Let $s(t)$ a data to be processed, $s_w(t)$ the whitened result of $s(t)$. This data is whitened as follows:
\begin{equation}
	s_w(t) = \mathscr{F}^{-1}[\frac{\tilde{s}(f)}{\sqrt{S_s(f)}}],
\end{equation}
where, $\tilde{s}(f)$ represents the Fourier transform of data $s(t)$, $\mathscr{F}^{-1}$ the inverse Fourier transform, and $S_s(f)$ the estimated power spectral density (PSD). In this work, the PSD is estimated using the Welch's average periodogram method \cite{welch1967use} with the implementation in Python library Matplotlib.

\textbf{Joint Time-Frequency Analysis~~~~}
This work proposes to do joint time-frequency analysis before detecting the existence of a GW signal. Wavelet packet transform is a typical time-frequency method \cite{blanco1998time,Book:Mallat:2009,Journal:Li:2015}. We chose the db1 of Daubechies wavelet basis \cite{Book:Daubechies:1992,Book:Mallat:2009} with 6-layer decomposition based on experimental confirmation. After wavelet packet decomposition, an observed strain with 65536 pixels is transformed into a  two-dimensional time-frequency map / a matrix with size $64\times 1024$.

\begin{figure}[!htbp]
	\centering
	\includegraphics[width=0.15\textwidth]{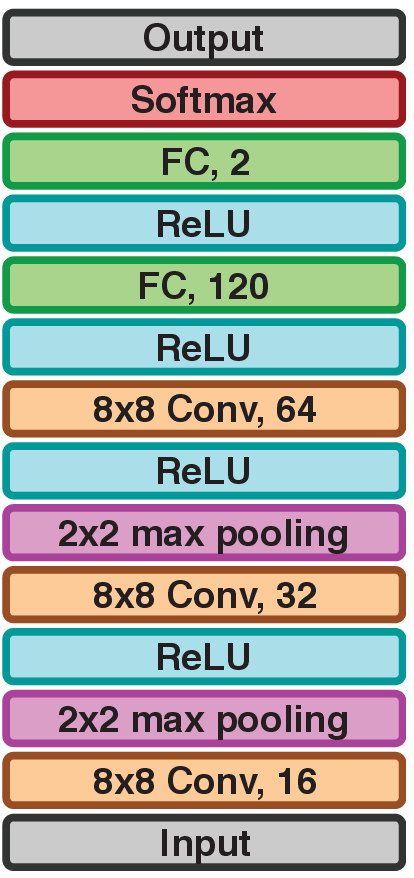}
	\caption{The architecture of the utilized CNN networks. This network consists of three convolutional layers (Convs) and two fully connected layers (FCs). All ``Conv" layers use a convolutional kernel with a configuration $8\times8$ and the numbers of kernels 16, 32, 64 on three ``Conv" layers according to their distances
		to the input layer from small to larger, respectively. The max pooling  is only performed on the two ``Conv" layers near the input layer, and the size of the corresponding kernel in pooling operation is $2\times2$. The output of the last layer is the probabilities that there exists any BNS signal in the input observation.}
	\label{fig:CNN_architecture}
\end{figure}

After the time-frequency analysis, the data are inputted into a convolutional neural network (CNN) to detect the existence of any BNS gravitational wave. CNN is a typical deep learning model based on local feature analysis, and has been widely used in image understanding, computer vision, and speech recognition, etc \cite{lecun2015deep}. This CNN consists of a series of basic computational units organized in a layer-by-layer style. The first layer accepts a two-dimensional time-frequency map as input, and the last layer gives the detection result of a CNN network (Fig. \ref{fig:CNN_architecture}). These two layers are also referred to input layer and output layer, respectively. This work takes the softmax function as the output layer \cite{goodfellow2016deep}, which estimates the probability of BNS gravitational wave existed in the input data.

The computational operations of a CNN network are mainly of three categories: convolutional operation, pooling operation and activation operation. The convolution operation is to extract features from the input image by computing a kind of similarity between a small square area of our input data and a convolutional kernel (CK). CK can be learnt from training data, and is a representation of the characteristics of the problem under dealing with, BNS gravitational wave detection. Therefore, convolution can preserves the spatial relationship between pixels. Typical Pooling operations are max pooling and average pooling. Based on the suggestions in \cite{li2017method}, this work adopts the max pooling. The max pooling is to split its input data into a series of disjoint rectangular subregions and calculate maximum values from each subregion.

Activation function is a nonlinear transformation used to enhance the performance of neural networks in complex problem, for example, the sigmoid, Rectified Linear Unit (ReLU), tanh and softmax \cite{goodfellow2016deep}. The softmax is used in output layer for classification/detection problem. In other nonlinear transformation layer, sigmoid function are one of the most widely used activation functions, ReLu is less computationally expensive than tanh and sigmoid. This work chooses the ReLU \cite{nair2010rectified} as the activation function in the inner nonlinear transformation layers.

In a CNN model, there exist some model parameters under being learnt from training data, for example, the weights between different computational units, the values of a matrix representing a CK, etc. The performance of these parameters are evaluated using cross entropy \cite{de2005tutorial}, and optimized by back propagation algorithm \cite{rumelhart1986learning}. The entire network is built on the open-source Tensorflow framework \cite{abadi2016tensorflow} with the stochastic optimization method ADAM \cite{kingma2014adam}. In a CNN network, furthermore, there are some parameters that could not be learnt from training data, but should specified experienced, for example, the number of convolution layers and fully connected layers, the size of convolutional kernel, et al. This kind parameters are referred to hyperparameters in literatures and determined based on a validation set.  And the architecture of the CNN model is shown in Fig. \ref{fig:CNN_architecture}.

This work uses two CNN models to deal with the data respectively from two detectors H1 and L1 and gives the final detection results by checking their coincidences of gravitational wave signal detection alert/trigger.

\section{Experimental Evaluations} \label{sec:experiments}

\noindent The proposed scheme is trained on theoretical data ``TrH1'' and ``TrL1'', %tested \& confirmed on the long stretches real BNS event GW170817,
and evaluated using theoretical data, ``TeH1'' and ``TeL1'',  on detection performance and efficiency with excellent results in being compared with the state-of-art method matched filtering. In matched filtering, the TBtr is utilized as a template bank. The TeH1 and TeL1 are used in testing matched filtering and the proposed scheme.

In the quantitative evaluations, this work uses receiver operator characteristics (ROC) curve. The ROC curve is created using sensitivity and false alarm rate (FAR), which is also referred to as false positive rate in literature. The sensitivity, also known as recall, true alarm rate or true positive rate, represents the fraction of the BNS gravitational wave that are successfully detected. The FAR is calculated as the ratio between the number of samples consisting of pure noises and disturbances (PND) wrongly detected as BNS gravitational wave signal and total number of actual PND samples. In application, we expect gravitational wave signal not to be missed and pure noise data not to cause false triggers. Therefore, the ROC curve of an appropriate detection scheme should approach the top-left corner of the FAR-sensitivity space.

\begin{figure}[!htbp]
	\centering
	\includegraphics[width=0.5\textwidth]{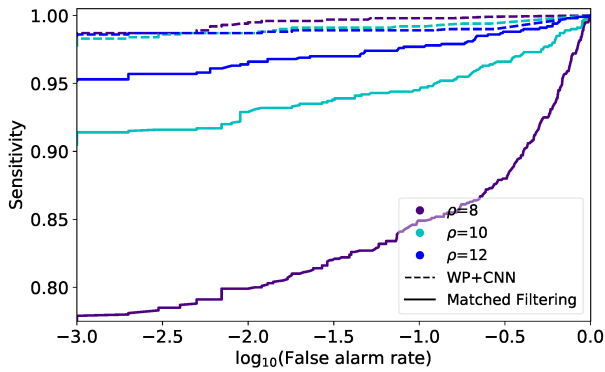}
	\caption{Quantitative performance evaluation of the proposed CNN-based scheme and its comparison with the matched filtering: Detection performance and its robustness to SNR. Two methods are tested on TeH1 and TeL1, the matched filtering are implemented using the template bank TBtr as knowledge-carrier. In these experimental results, the CNN-based scheme shows strong robustness to FAR and evident superiority on sensitivity and false alarm rate. In this figure, the $\rho$ is the optimal matched-filter SNR defined in equation (\ref{Equ:rho:SNR}).}
	\label{fig:roc}
\end{figure}

\textbf{Quantitative performance evaluation and comparisons~~} The corresponding results are presented in Fig. \ref{fig:roc}. In this figure, three dashed curves are on top and the tested results of the proposed scheme, the bottom three solid curves are that of the matched filtering method. Furthermore, the sensitivity of CNN-based scheme shows strong robustness to FAR and are higher than 97.2\% in case of FAR$\geq 0.001$.  Therefore, the above experiments show the potentials of the CNN-based method in detecting gravitational waves of binary neutron stars.

\begin{figure}[!htbp]
	\centering
	\includegraphics[width=0.5\textwidth]{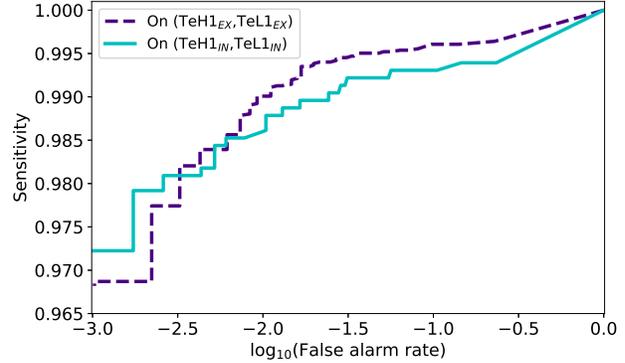}
	\caption{Generalization. The parameter configuration of a test sample in `TeH1$_{IN}$' and `TeL1$_{IN}$' is in the range which generated training samples, while that in `TeH1$_{EX}$' and `TeL1$_{EX}$' is out of the range. }
	\label{fig:inside_outside_roc}
\end{figure}

\textbf{Generalization~~} In application, it's impossible to guarantee that the signal to be detected shares the same parameter values with our experiences. In this work, the `experiences' refers to the templates TBtr used in producing training samples ``TrH1'' and ``TrL1'' and the `parameters' are the masses $m_1$ and $m_2$ of a BNS system. It is important to test the performance of a proposed scheme on some samples  novel compared with our experience, and this problem is called generalization in machine learning community. Based on this kind consideration, this work independently generated two template banks of noise-free BNS gravitational wave signals, ``TBtr" and ``TBte", where ``TBte" is used for producing test sample sets ``TeH1'' and ``TeL1''. Therefore, the experimental results in Fig. \ref{fig:inside_outside_roc} show that the proposed scheme have excellent generalization capability.

As for generalization, two kinds of cases can be distinguished based on whether the parameter configuration of a test sample is in the range that the experiences TBtr are sampled from. Therefore, the test sets `TeH1' and `TeL1' can be further divided into `TeH1$_{IN}$', `TeL1$_{IN}$', `TeH1$_{EX}$' and `TeL1$_{EX}$'.  The `$IN$' and `$EX$' are the abbreviations of interpolation and extrapolation, respectively. Experiments didn't show any explicit influence on performance from these two cases (Fig. \ref{fig:inside_outside_roc}).

\begin{figure}[!htbp]
	\centering
	\includegraphics[width=0.5\textwidth]{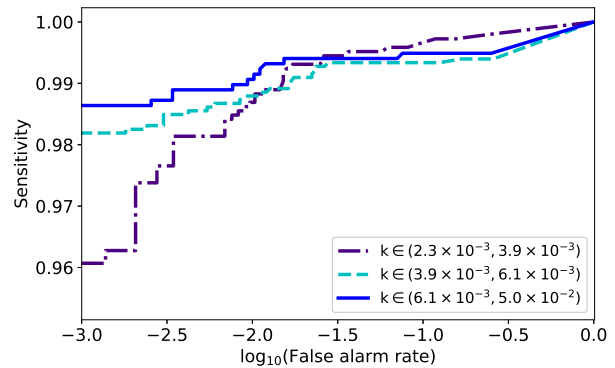}
	\caption{Performance evaluation of the proposed CNN-based scheme using test sets with different $k$ in Eq. (\ref{eq:hn}). This evaluation is conducted on union of the two test sets TeH1 and TeL1.}
	\label{fig:k_roc}
\end{figure}

\textbf{Performance related with $k$ ~~} As being described in section \ref{sec:data}, the experimental data with various SNR are computed by adjusting the parameter $k$ in equation (\ref{eq:hn}). Therefore, one problem is how about the relationship between performance of the proposed CNN-based scheme and $k$.  Experimental results in Fig. \ref{fig:k_roc} show that there exist some positive correlations between the detection performance and $k$. One potential usage of this relationship is to theoretically estimate the possible detection performance on some observations in case of the corresponding $k$ being known or being estimated using some kind methods. Considering that the focus of this work is the detection scheme, therefore, we postpone the specific performance estimation from $k$ to future work.

\textbf{Efficiency~~} One typical characteristic of the proposed scheme is its efficiency. In the experiments of this work, the WP-CNN scheme takes approximately 0.027 seconds to process a sample on a computer with an Intel(R) Xeon(R) CPU E7- 4830, 2.13GHz processor, while matched filtering about 261 seconds for a sample.

\section{Conclusions}

\noindent This work proposed a BNS gravitational wave detection scheme based on wavelet packet (WP) transform and convolutional neural network (CNN), which is evaluated on theoretical data. Quantitative evaluation show that the proposed CNN-based scheme has evident superiority on detection sensitivity. On efficiency, the current experiments show that the CNN-based scheme is more than 960 times faster than the matched filtering. In this work, the CNN-based scheme and matched filtering are all implemented based on one-core computation and their parallel optimization are postponed to future work.

\acknowledgements{Authors are grateful for valuable suggestions from Xilong Fan, and thank the supports by the National Natural Science Foundation of China (grant No: 11973022, 61273248, 61075033), the Natural Science Foundation of Guangdong Province (2014A030313425, S2011010003348), China Scholarship Council (201706755006), and the Joint Research Fund in Astronomy (U1531242) under cooperative agreement between the National Natural Science Foundation of China (NSFC) and Chinese Academy of Sciences (CAS), and the Major projects of the joint fund of Guangdong and the National Natural Science Foundation (U1811464) to our free explorations.}

%\bibliographystyle{fop}
%\bibliography{library,books}

%%%%%%%%%%%%%%%%%%%%%%%%%%%%%%%%%%%%参考文献的排法%%%%%%%%%%%%%%%%%%%%%%%%%%%%%%%%%%%%
%%%%%%%%%%%%%%%%%%%%%%%%%%%%%%%%%%%%注：\href{}{}%%%第一个括号表示需要链接的网址,第二个花括号表示锭接网址的内容%%%%%%%%%%%%%%%%
\raggedend
%\input{ref.bbl}
%\begin{multicols}{2}

\begin{small}

\end{small}

%\end{multicols}

\end{document}